\documentclass[a4paper]{article}
\usepackage{Odyssey2024}
\usepackage{epsfig,amssymb,amsmath,balance,cite}
\ninept

\title{Device Feature based on Graph Fourier Transformation with Logarithmic Processing For Detection of Replay Speech Attacks}

\makeatletter
\def\name#1{\gdef\@name{#1\\}}
\makeatother
\name{\em Mingrui He$^{\star}$, Longting Xu$^{\star}$, Han Wang$^{\star}$, Mingjun Zhang$^{\star}$ and Rohan Kumar Das$^{\dagger}$}
\address{$^{\star}$College of Information Science and Technology, Donghua University, China\\
    $^{\dagger}$Fortemedia Singapore, Singapore\\
    {\small \tt xlt@dhu.edu.cn}}
\usepackage{multirow}
\usepackage{booktabs}
\usepackage{graphicx}
\usepackage{float}
\usepackage{subfig}
\begin{document}
\maketitle

\begin{abstract}

The most common spoofing attacks on automatic speaker verification systems are replay speech attacks. Detection of replay speech heavily relies on replay configuration information. Previous studies have shown that graph Fourier transform-derived features can effectively detect replay speech but ignore device and environmental noise effects. In this work, we propose a new feature, the graph frequency device cepstral coefficient, derived from the graph frequency domain using a device-related linear transformation. We also introduce two novel representations: graph frequency logarithmic coefficient and graph frequency logarithmic device coefficient. We evaluate our methods using traditional Gaussian mixture model and light convolutional neural network systems as classifiers. On the ASVspoof 2017 V2, ASVspoof 2019 physical access, and ASVspoof 2021 physical access datasets, our proposed features outperform known front-ends, demonstrating their effectiveness for replay speech detection.
\end{abstract}

\section{Introduction}
Automatic speaker verification (ASV) systems are now commonly used in various application scenarios that leverage the differences in vocal tract shape, tone, and pronunciation habits of speakers to recognize them by comparing the target utterance with the registered one~\cite{RN83}. While the emergence of deep learning and artificial neural networks have greatly improved the recognition ability of ASV systems, these systems face threats from spoofing attacks~\cite{spoof_review, bib:Attacker_overview2020}. 

ASV systems encounter four primary types of spoofing attacks, namely, impersonation, text-to-speech synthesis (TTS), voice conversion (VC), and replay attacks~\cite{spoof_review}. Among these, TTS and VC are logical access attacks that require the attackers to know about TTS and VC systems to produce the target speaker's speech~\cite{vcc2020summary,vcc2020objective}. Impersonation is found to show comparatively less threat to ASV systems as it mainly tries to mimic the behavioral characteristics of the target speaker~\cite{mimicking}. On the other hand, replay attacks are much easiser to produce by using recorded samples of the target speaker and also show a very high threat to break ASV systems~\cite{Li2016_spoof_TD}. In this work, we focus on the detection of replay speech attacks.  

The replay speech contains the recording as well as playback device information and the characteristics of the background environment. It is critical to capture such information for the detection of replay attacks as they contain strong traits of the target speaker. In this regard, we focus on exploring novel signal representations that can help to expose the device information. Some of the strong feature representations that have been explored previously include constant-Q cepstral coefficients (CQCC)~\cite{CQCC_odyssey2016,CQCC_CSL}, linear frequency cepstral coefficients (LFCC), cochlear filter cepstral coefficient and instantaneous frequency~\cite{PatelINTERSPEECH2015}. However, these feature representations did not consider utilizing any device information explicitly. 

In general, a replay attack involves recording a registered sound in a specific environment and subsequently replaying it using speakers to deceive an ASV system. The difference between genuine speech and replayed speech is mainly due to the physical characteristics of the recording device, loudspeaker, and ambient noise. This creates a filter in the frequency domain that affects the intrinsic differences between the two types of speech. Hence, in our recent work, we proposed a novel device-related linear transformation strategy to separate non-device information from replay speech by using the corresponding genuine speech to emphasize device information in the replay speech~\cite{Longting_deviceFeat}. This is based on the rationale that the difference between genuine speech and replayed speech is mainly due to the physical characteristics of the recording device, loudspeaker, and ambient noise~\cite{Jichen_deviceFeat}. We then developed three novel device features based on this device-related linear transformation strategy, which were obtained from the octave domain by combining constant-Q transform and octave subband transform, the linear domain on LFCC, and the mel domain on mel frequency cepstral coefficient (MFCC). 

Graph signal processing (GSP) is another emerging method that has been useful for several applications including speech processing. It constructs the graph topology by the potential relationship between speech samples and then utilizes graph Fourier transform (GFT) to transform speech signal from the graph domain to the graph frequency domain for reflecting the structural relationship between speech samples. We utilized it to derive a feature namely, graph frequency cepstral coefficient (GFCC) in \cite{GFCC_odyssey}. Although GFCC captures more hidden information in speech and emerged as a promising front-end compared to existing front-ends, it did not consider any specific device-related information to handle replay speech attacks. 

\begin{figure*}[t]
 \centering
 \subfloat[]{\includegraphics[width=1.5in]{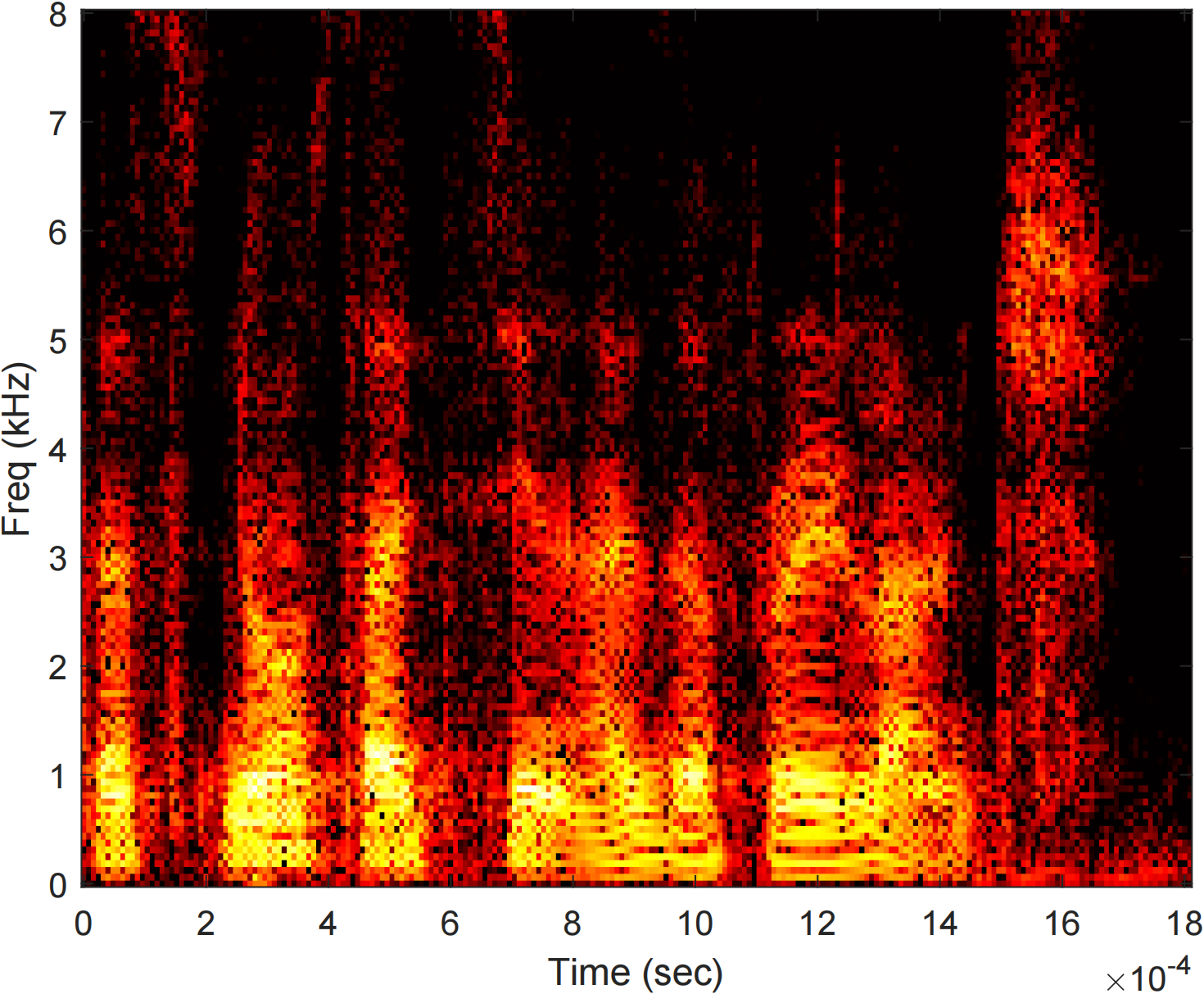}\label{fig:gfcc genunie Spectrograms}}
 \hfil
 \subfloat[]{\includegraphics[width=1.5in]{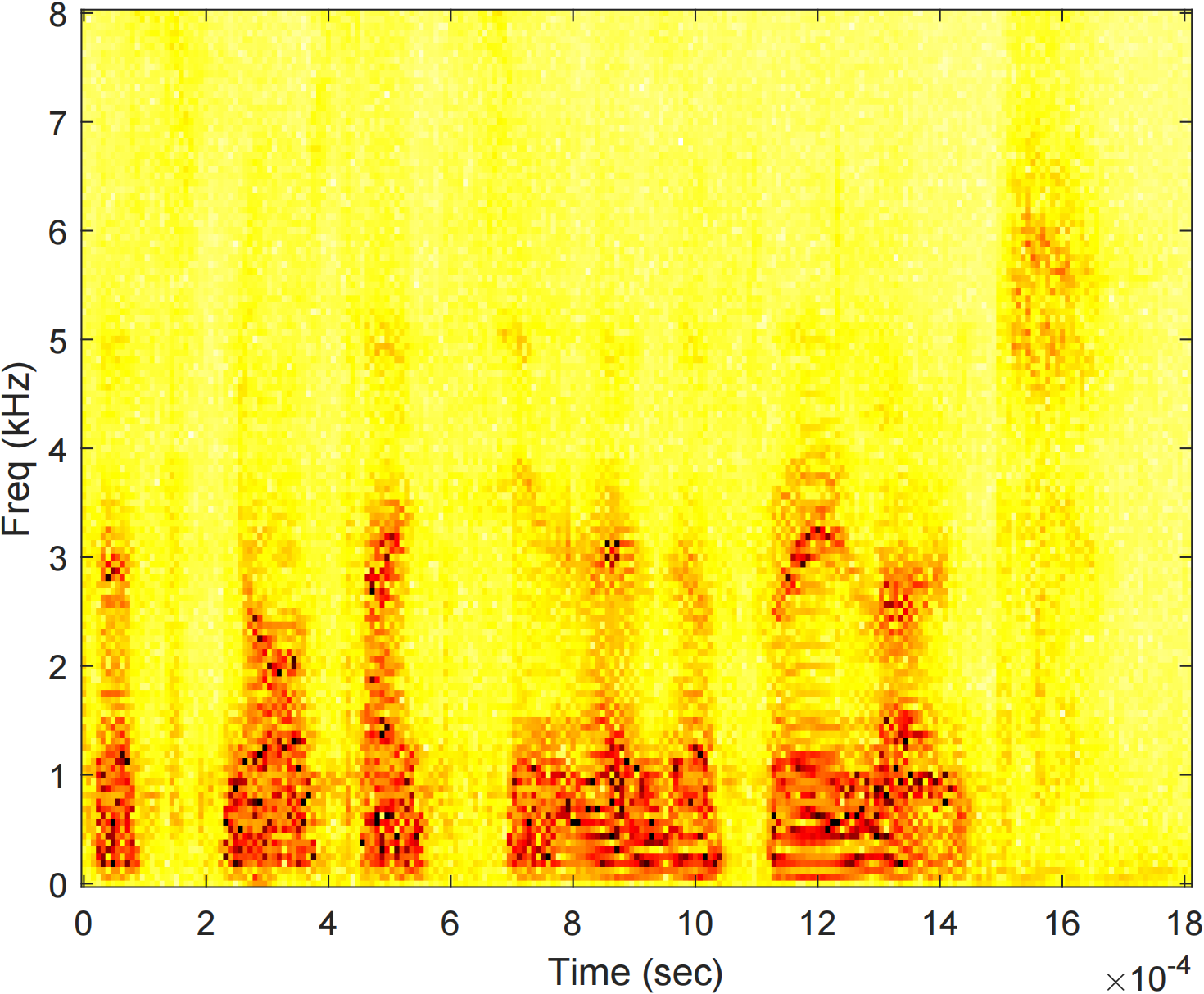}\label{fig:GFLC genunie Spectrograms}}
 \hfil
 \subfloat[]{\includegraphics[width=1.5in]{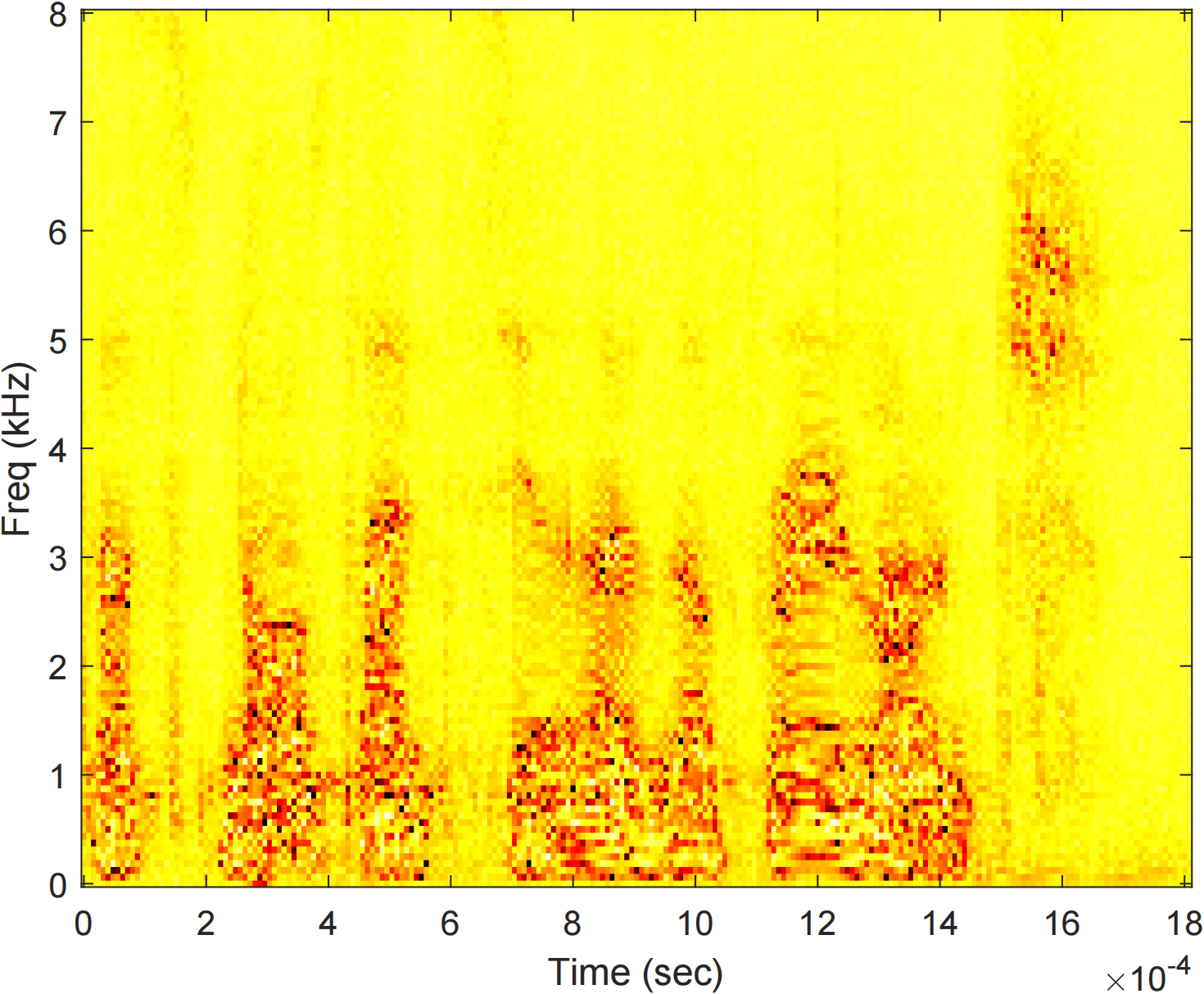}}
 \hfil
 \subfloat[]{\includegraphics[width=1.5in]{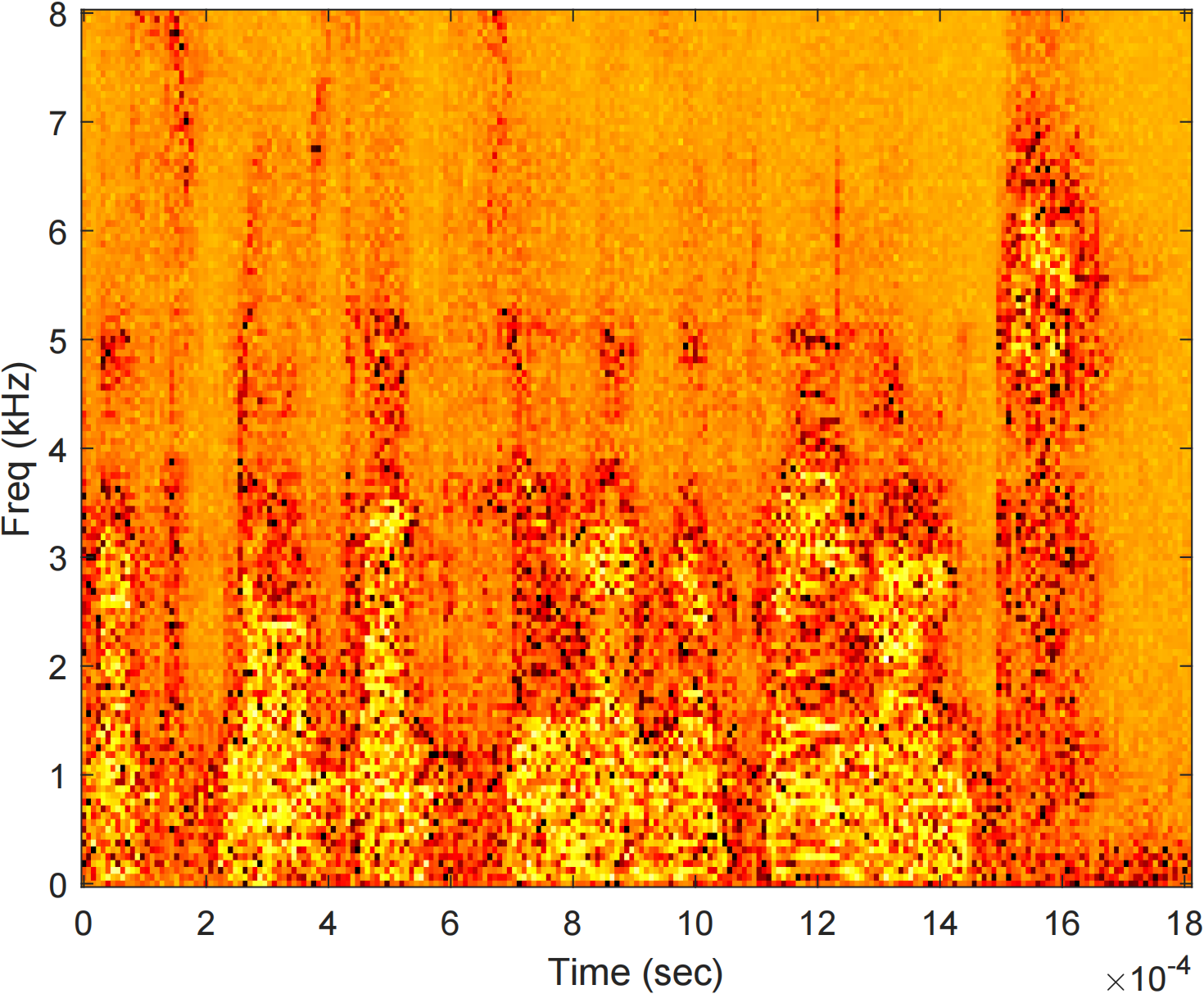}}
 \caption{Spectrograms of a genuine speech, using different numbers of logarithmic operations (a) original spectrogram (GFCC). (b) spectrogram after single logarithimic processing (GFLC). (c) spectrogram after twice logarithmic processing. (d) spectrogram after thrice logarithmic processing.}
 \label{fig:genunie Spectrograms}
\end{figure*}

\begin{figure*}[t]
 \centering
 \subfloat[]{\includegraphics[width=1.5in]{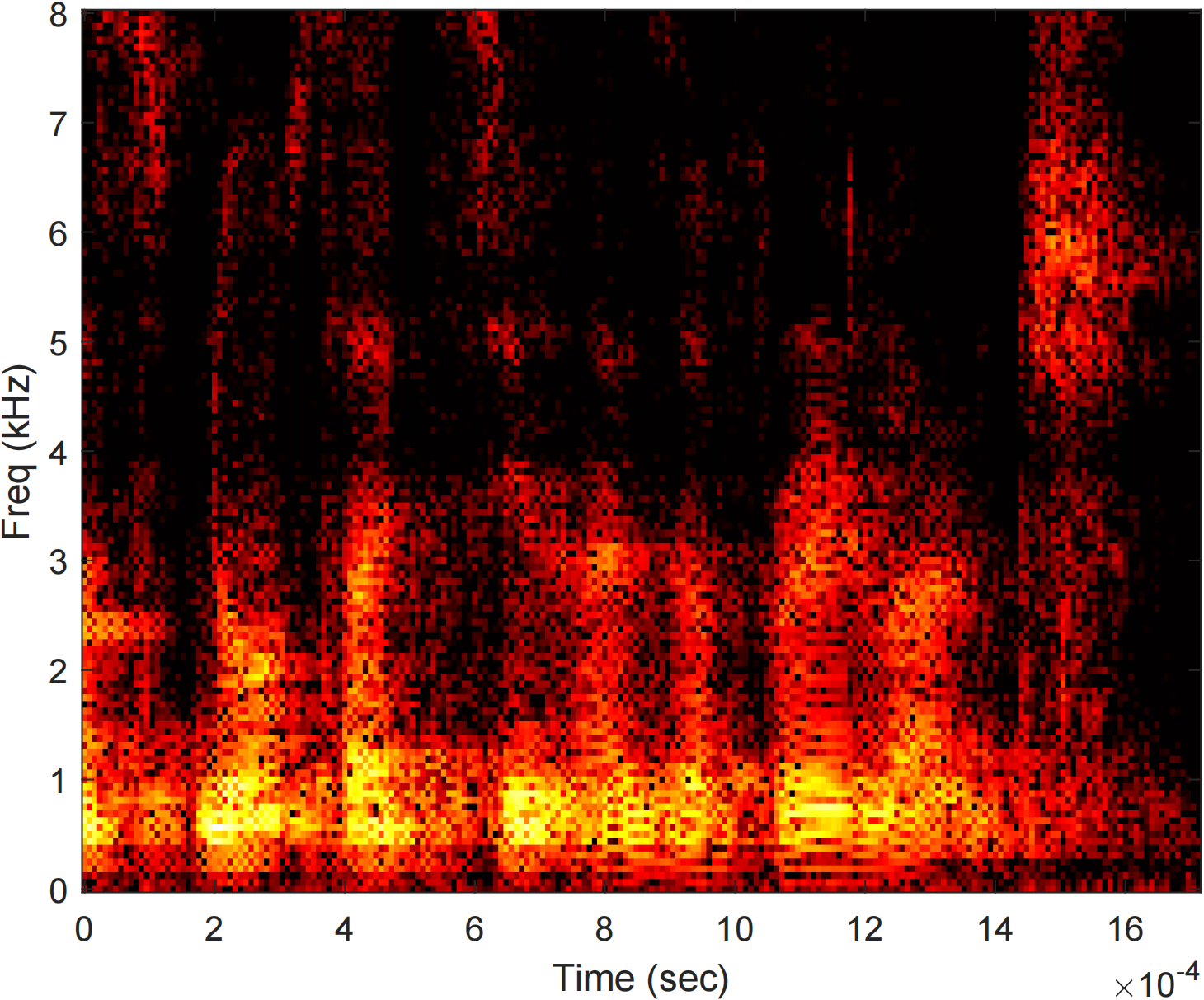}\label{fig:gfcc spoofed Spectrograms}}
 \hfil
 \subfloat[]{\includegraphics[width=1.5in]{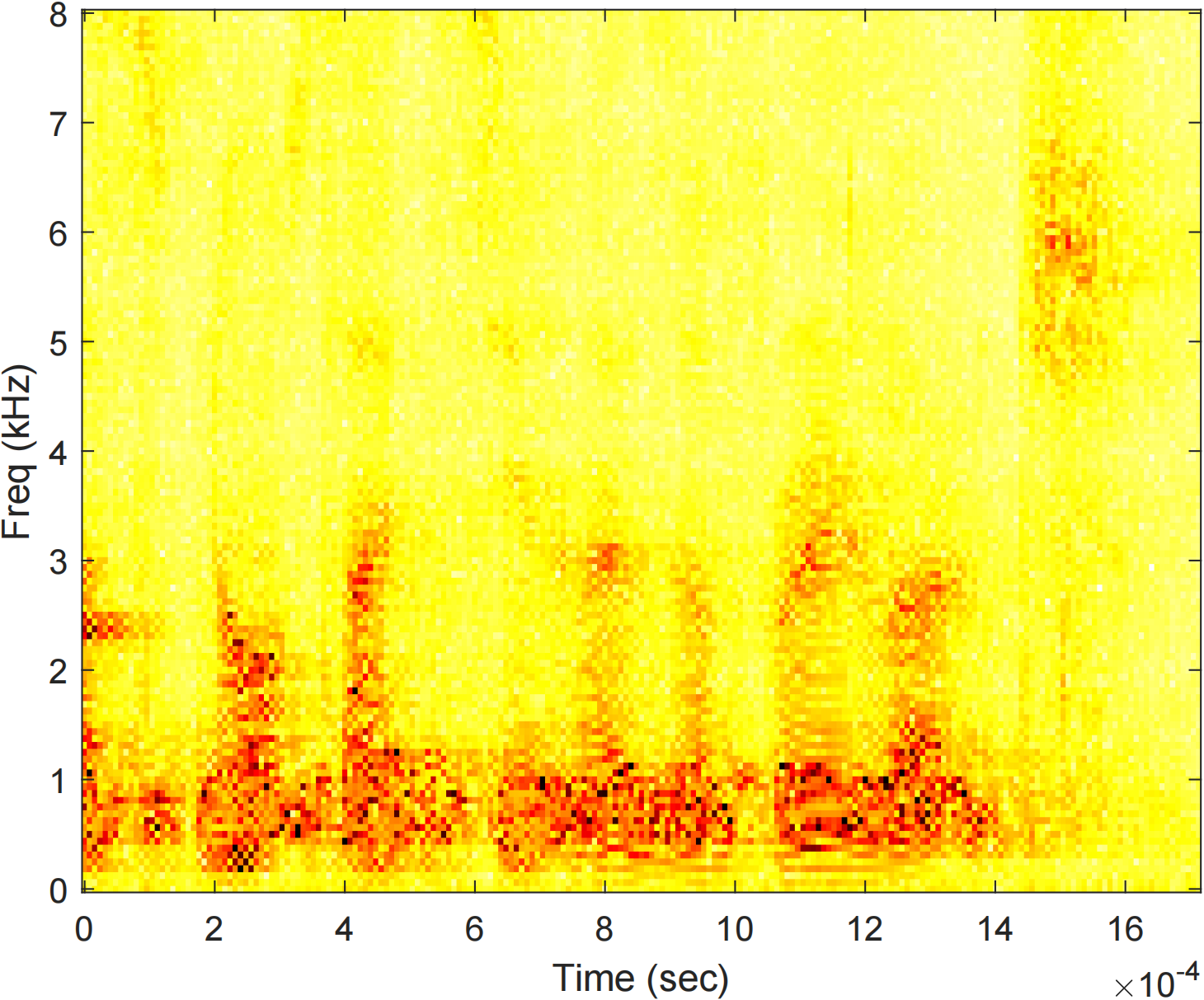}\label{fig:GFLC spoofed Spectrograms}}
 \hfil
 \subfloat[]{\includegraphics[width=1.5in]{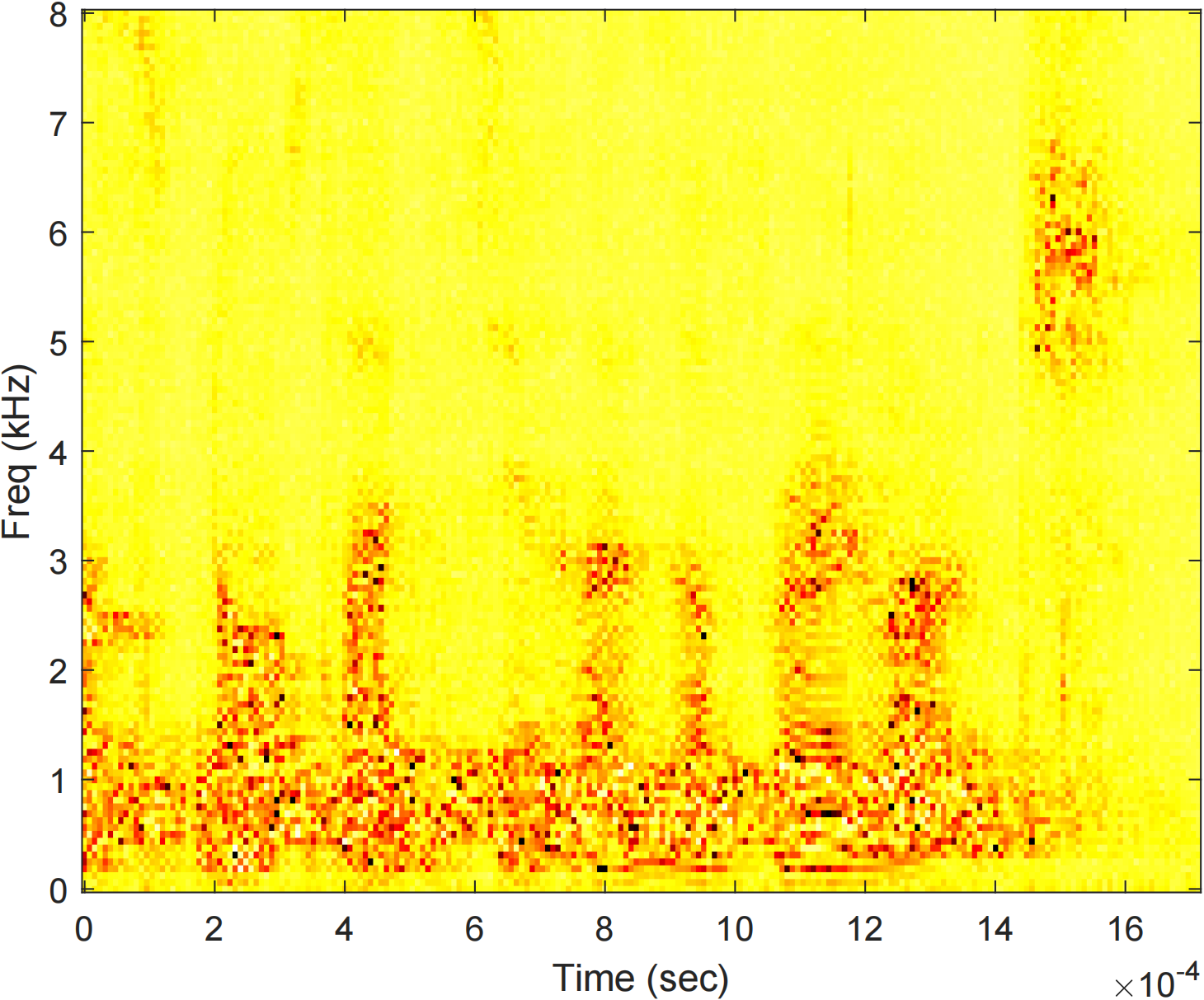}}
 \hfil
 \subfloat[]{\includegraphics[width=1.5in]{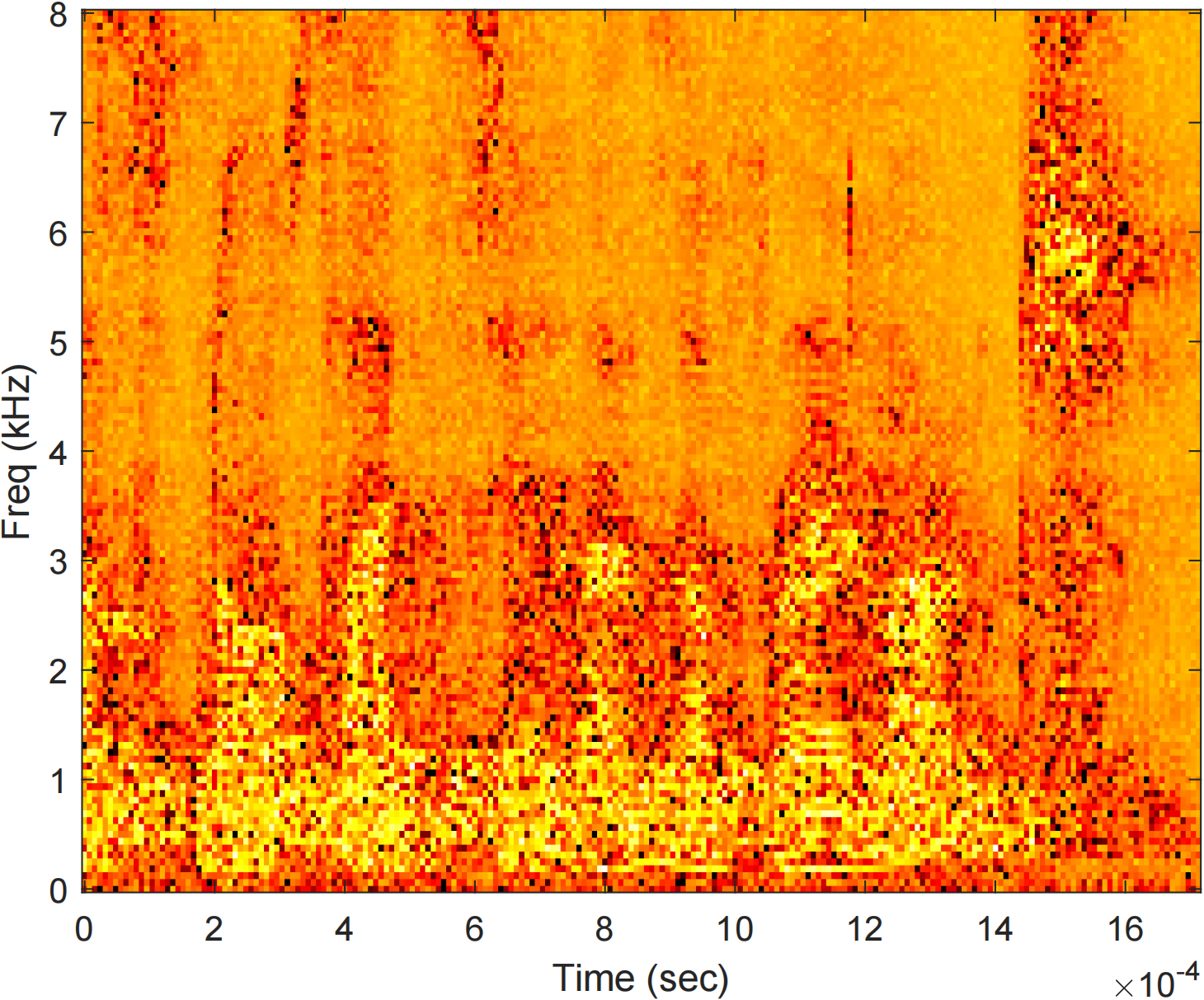}}
 \caption{Spectrograms of a spoofed speech, using different numbers of logarithmic operations (a) original spectrogram (GFCC). (b) spectrogram after single logarithmic processing (GFLC). (c) spectrogram after twice logarithmic processing. (d) spectrogram after thrice logarithmic processing.}
 \label{fig:spoofed Spectrograms}
\end{figure*}

In this work, we use the device-related linear transformation on the GFCC feature to propose a novel feature referred to as graph frequency device cepstral coefficient (GFDCC) that is expected to benefit from both explicit device-related information along with the GSP technology. Moreover, spectrogram analysis following different times of using logarithmic processing on the features revealed more pronounced distinctions between original and single logarithmic processed spectrograms of genunie and spoofed speech, as illustrated in Figure~\ref{fig:genunie Spectrograms} and Figure~\ref{fig:spoofed Spectrograms}. Consequently, we applied logarithmic operations to the features after GFT and introduced two novel representations, namely,  graph frequency logarithmic coefficient (GFLC) and graph frequency logarithmic device coefficient (GFLDC) based on GFCC and GFDCC. We study these features with two different classifiers Gaussian mixture model (GMM) and light convolutional neural network (LCNN) for replay speech detection on ASVspoof 2017 V2, ASVspoof 2019 physical access and ASVspoof 2021 physical access datasets~\cite{ASVspoof2017,ASVspoofV2,ASVsppof2019_paper,ASVspoof2021_paper}.

The rest of this paper is organized as follows.   Section~\ref{section:Logarithmic processing in GFLC} describes the proposed feature GFLC with logarithmic processing, and Section~\ref{section:Device Information in GFDCC} presents the proposed GFDCC and GFLDC with device-related linear transformation. Section~\ref{section:Experiments} introduces the datasets used in the experiment and the detailed experimental setup.  The experimental results and analysis on replay speech detection are given in Section~\ref{section:Results and Analysis}. Finally, the work is concluded in Section~\ref{section:Conclusions}.

\section{Logarithmic processing in GFLC}
\label{section:Logarithmic processing in GFLC}
\subsection{GFCC feature}

First, we would like to briefly discuss about GFCC feature that we proposed earlier in~\cite{GFCC_odyssey}. It is a short-term processing based feature derived using GFT after constructing the graph signal and finally applying discrete cosine transform (DCT) to the log power spectrum to compute the cepstral coefficients, as given by:
\begin{align}
   GFCC(z) &= c(z) \sum_{i=0}^{N-1} \log|\hat{y}_i|^2 \cos{[\frac{(i+0.5)\pi}{N}z]}
\end{align}
where \(\log|\hat{y}_i|^2\) is the log power spectrum of the speech signal, \(N\) is the number of points of the speech signal, and \(c(z)\) can be considered as a compensation coefficient, which is defined as:
\begin{align}
c(z) =
\begin{cases}
\sqrt{\frac{1}{N}}, & \mbox{if }z\mbox{ = 0} \\
\sqrt{\frac{2}{N}}, & \mbox{else }
\end{cases}
, z\mbox{ = 0 , 1, \dots , N-1}
\end{align}

\subsection{Proposed GFLC feature with logarithmic processing}
Figure~\ref{fig:genunie Spectrograms} and Figure~\ref{fig:spoofed Spectrograms} illustrates that both the original spectrogram and the single logarithmic processed spectrogram are more effective in discriminating between genuine speech and replay speech. Among these, Figure~\ref{fig:gfcc genunie Spectrograms} and Figure~\ref{fig:gfcc spoofed Spectrograms} represent the genuine and spoofed speech spectrograms of GFCC, while Figure~\ref{fig:GFLC genunie Spectrograms} and Figure~\ref{fig:GFLC spoofed Spectrograms} depict the genuine and spoofed speech spectrograms of proposed GFLC, respectively. In comparison to Figure~\ref{fig:gfcc genunie Spectrograms} and Figure~\ref{fig:gfcc spoofed Spectrograms} and the other four subgraphs, Figure~\ref{fig:GFLC genunie Spectrograms} and Figure~\ref{fig:GFLC spoofed Spectrograms} exhibit more pronounced distinctions and also present clearer spectrograms. Hence, we apply logarithmic processing to the features after GFT, followed by vertical splicing with the original GFT features to obtain GFLC, defined as:
\begin{align}
   GFLC(z) &= c(z) \sum_{\omega=0}^{N-1} \log|[\hat{y}_i;ln(|\hat{y}_i|)]|^2 \cos{[\frac{(i+0.5)\pi}{N}z]}
\end{align}
where the semicolon indicates that concatenate the original and log-compressed magnitudes in order to extract the real-fake information from both amplitudes simultaneously. The rest of the extraction steps of GFLC are consistent with those of GFCC.

\begin{figure}[t]
\centering
\subfloat[GFLC]{\includegraphics[width=1.5in]{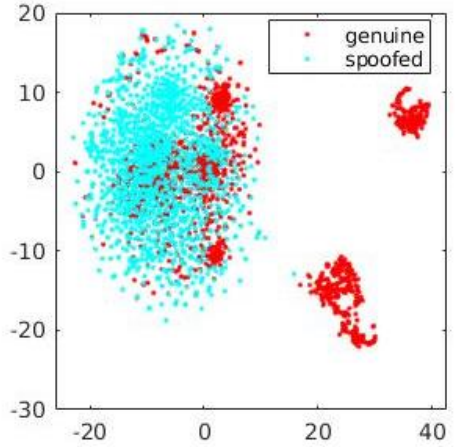}}%
\label{fig:tsne 1}
\hfil
\subfloat[GFCC]{\includegraphics[width=1.5in]{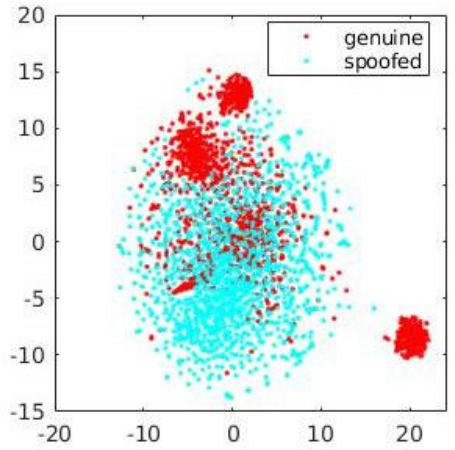}}%
\label{fig:tsne 2}
\caption{t-SNE visualizations: (a) GFLC (b) GFCC.}
\label{fig:tsne}
\end{figure}

Moreover, the dimension-reduced embedding visualization is shown in Figure~\ref{fig:tsne}. The same t-distributed Stochastic Neighbor Embedding (t-SNE) projections are applied to GFLC and GFCC on training set of ASVspoof 2017 Version 2.0 corpus. It is found that the inter-class distance of GFLC, that is, the distance between genuine speech and spoofed speech sample points, is significantly greater than that of GFCC.

\begin{figure*}[t]
  \centering
  \includegraphics[width=\linewidth]{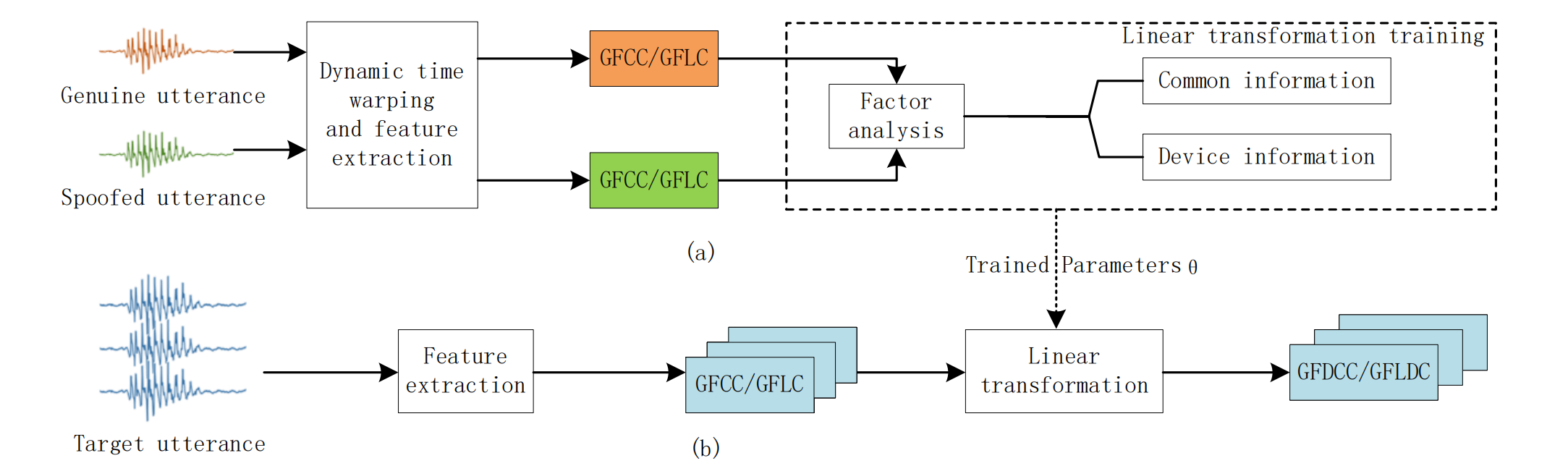}
  \caption{An overview of incorporating device information in GFDCC and GFLDC 
  features: (a) Training the linear transformation parameters \(\theta\) based on ASVspoof2017 training set after DTW (b) GFDCC and GFLDC extraction on original datasets when GFCC and GFLC are used as the input to the trained linear transformation.}
  \label{fig:overview}
\end{figure*}

\section{Device Information in GFDCC and GFLDC}
\label{section:Device Information in GFDCC}

\subsection{Device-related linear transformation}
\label{section:Device-related linear transformation}

As we discussed in the introduction section, the replay speech contains strong target speaker characteristics that include the speaker's speech, emotion, intonation and other common information. In addition, the replay speech also contains the physical characteristics of the recording and playback devices as well as the environmental noise during recording. However, different genuine-replay pairs of speech may share similar components, such as physical characteristics and environment of the recording device used for genuine speech, despite differences in speaker characteristics and device information during the playback process~\cite{Longting_deviceFeat}. 

Under this premise, if we consider that \(\phi_{s,l,r}\) and \(\phi_{g,l,r}\) represent the \(r\)-th frame vector of input feature sequences of the \(l\)-th pair of replay speech utterance and corresponding genuine speech utterance, respectively, then the parallel factor analysis can be expressed as: 
\begin{align}
   \phi_{s,l,r} &= \mu + \mathbf{F}h_l + \epsilon_{s,l,r}  \\
   \phi_{g,l,r} &= \mu + \mathbf{F}h_l + \epsilon_{g,l,r} 
\end{align}
where \(\mu\) represents the universal mean vector of all speech utterances, \(\mathbf{F}\) denotes the projection matrix of a linear transformation, and \(h_l\) is the common factor vector, which contains the same components of speaker, speech content, and speaker emotion of the \(l\)-th parallel replay utterance and its corresponding genuine utterance. The residual vectors \(\epsilon_{s,l,r}\) and \(\epsilon_{g,l,r}\) have a Gaussian distribution with zero-mean and variance \(\Sigma\) for replay speech and genuine speech, respectively, and they imply that the uncommon variability in the utterances is more relevant to the device factor and recording environment component.

Considering \(k_{s,l}\) as the number of frames of the {\it l}-th replay speech utterance feature sequence, while the frame number of feature sequences of the corresponding genuine speech utterance is represented by \(k_{g,l}\). The matrix of the {\it l}-th parallel speech utterance with the same observation \(h_l\) can be obtained as follows:
\begin{align}
    \begin{bmatrix}
    \phi_{s,l,1}  \\
    \vdots\\
    \phi_{s,l,k_{s,l}} \\
    \phi_{g,l,1} \\
    \vdots\\
    \phi_{g,l,k_{g,l}}
    \end{bmatrix}
    &=
    \begin{bmatrix}
    \mu  \\
    \vdots\\
    \mu \\
    \mu \\
    \vdots\\
    \mu
    \end{bmatrix}
    +
    \begin{bmatrix}
    \mathbf{F}  \\
    \vdots\\
    \mathbf{F} \\
    \mathbf{F} \\
    \vdots\\
    \mathbf{F}
    \end{bmatrix}
    h_l
    +
    \begin{bmatrix}
    \epsilon_{s,l,1}  \\
    \vdots\\
    \epsilon_{s,l,k_{s,l}} \\
    \epsilon_{g,l,1} \\
    \vdots\\
    \epsilon_{g,l,k_{g,l}}
    \end{bmatrix}
\end{align}

The matrix  \(\hat{\phi}_l\) of the \(l\)-th parallel speech utterance with the same observation \(h_l\) can be obtained as follows:
\begin{align}
   \hat{\phi}_l &= \hat{\mu} + \mathbf{\hat{A}}h_l + \hat{\epsilon}_l
\end{align}
where \(\hat{\mu}\) and \(\mathbf{\hat{A}}\) are the result of the superposition of \(\mu\) and \(\mathbf{F}\) respectively, \(h_l \sim \mathcal{N} (0,\mathbf{I})\), and the residual part \(\hat{\epsilon}_l \sim \mathcal{N} (0,\hat{\Sigma})\).

Then the parameters \(\theta = \{\mu,\mathbf{F},\Sigma\}\) are updated using the EM algorithm, which involves the E step and the M step, and the device feature is extracted based on the linear transformation model  \(\theta\). The posterior mean of \(h_l\) is deduced at step E
\begin{align}
   E[h_l] &= \mathbf{\hat{L}}^{-1}\mathbf{\hat{A}}^T\mathbf{\hat{\Sigma}}^{-1}(\hat{\phi}_l - \hat{\mu})
\end{align}
where \(\mathbf{\hat{L}}^{-1}\) is obtained by
\begin{align}
   \mathbf{\hat{L}}^{-1} &= (\mathbf{I} + \mathbf{\hat{A}}^T\mathbf{\hat{\Sigma}}^{-1}\mathbf{\hat{A}})^{-1}
\end{align}
The universal mean vector \(\mu\) used here can be obtained by training over the entire training dataset. The factor analysis approach used here is inspired from that used in case of i-vector modeling~\cite{Dehak2011}.

\subsection{Dynamic time warping}
Let $G = [g_1,g_2,...g_m]$ be a genuine utterance series and $S = [s_1,s_2,...s_n]$ be its corresponding spoofed utterance series. Dynamic time warping (DTW) tries to find the best alignment between $G$ and $S$ such that the accumulated difference between the mapping points is minimum. During mapping, it is possible that multiple points in one series map to one point in the other series. DTW finds such a mapping by building a cost matrix $C$ between each pair of points in $G$ and $S$ where each element \(d(i, j)\) of the matrix is the cost of aligning $g_i$ with $s_j$. DTW then uses dynamic programming, shown as the following recursive formula, to minimize the sum of the elements on a path from the bottom left to the top right of the cost matrix. The path is known as the warping path; at each element of the cost matrix, the path goes right, up, or up right \cite{shen2021tcdtw}.
\begin{align}
   DTW(i,j) &= d(i,j) + min 
\begin{cases}
DTW(i-1,j) \\
DTW(i,j-1) \\
DTW(i-1,j-1)
\end{cases}
\end{align}

\subsection{Proposed GFDCC and GFLDC features with device information}

Fig.~\ref{fig:overview} shows the overview of GFDCC and GFLDC feature extraction by incorporating device information. First, it requires some parallel training data (genuine and corresponding replay speech) on which DTW is applied to obtain equal length feature matrices of GFCC or GFLC features. In this work, we used ASVspoof 2017 V2 training set to form the parallel data. Once equal length GFCC or GFLC features are extracted on the parallel data, a factor analysis approach is applied to separate the common information and device characteristics as shown in Fig.~\ref{fig:overview} (a). Then, the hyperparameters \(\theta\) of device-related linear transformation described in Section~\ref{section:Device-related linear transformation} are extracted from the parallel data, by training with the EM algorithm. Finally, considering linear convolution effects become additive, the GFCC or GFLC of target utterances from the original datasets are used as input for the trained device-related linear transformation \(\theta\) as shown in Fig.~\ref{fig:overview} (b), and the resulting output are used to obtain the proposed GFDCC and GFLDC features containing device information, given by:
\begin{align}
   GFDCC(z) &= GFCC(z) - \mu - \mathbf{F}{E}[h] \\
   GFLDC(z) &= GFLC(z) - \mu - \mathbf{F}{E}[h] 
\end{align}
where \(E[h]\) represents the mean value of the common factor vector \(h_l\) obtained by EM algorithm.

\section{Experiments}
\label{section:Experiments}
\subsection{Datasets}
The studies in this work are carried on ASVspoof2017 V2 , ASVspoof 2019 PA and ASVspoof 2021 PA datasets~\cite{ASVspoofV2,ASVspoof_database_CSL,ASVspoof2021_paper}. The genuine utterances for ASVspoof 2017 V2 corpus were taken from the RedDots database, while the same for ASVspoof 2019 PA and ASVspoof 2021 PA were taken from the VCTK database. It is also noted that ASVspoof 2017 V2  and ASVspoof 2021 PA are realistic replay databases, whereas ASVspoof 2019 PA considers replay attacks in simulated settings. We used the ASVspoof 2017 V2 training set after DTW for device-related linear transformation parameters extraction, and the original evaluation sets of ASVspoof 2017 V2,  ASVspoof 2019 PA and ASVspoof 2021 PA to evaluate the system performance.  We would also like to note here that we used ASVspoof 2017 V2 and ASVspoof 2019 PA datasets, respectively, when evaluated on ASVspoof2021 PA. Since, ASVspoof 2017 V2 and  ASVspoof 2021 PA are both realistic replay speech database, while the training set under its evaluation protocol only considers the training set from ASVspoof 2019 PA, which is a simulated database; hence, creating a mismatched scenario that has other criticality for benchmarking. 

The details of the three datasets are shown in Table~\ref{tab:datasets}. In addition to evaluation set, ASVspoof 2021 PA provides progress set, where the data is utterance-disjoint, and two hidden attack sets, where the non-speech intervals are removed, with the motivation of evaluating performance when the spoofed speech detector is restricted to operating only on speech segments~\cite{asvspoof2021database}.

The study evaluates the proposed method using two performance metrics, namely equal error rate (EER) and tandem detection cost function (t-DCF) \cite{tDCF_Tomi_IEEE}. ASVspoof 2017 V2 evaluation protocol considers EER as the only metric, while ASVspoof 2019 PA and ASVspoof 2021 PA consider it as a secondary metric along with t-DCF as the primary measure. 

\begin{table}[t]
  \caption{A summary of ASVspoof2017 V2, ASVspoof2019 physical access and ASVspoof 2021 physical access datasets.}
  \label{tab:datasets}
  \resizebox{\linewidth}{!}
{
  \centering
  \begin{tabular}{|c|c|c|c|c|}
    \hline
    \textbf{Database}
                    &\textbf{Set}&\textbf{Genuine}&\textbf{Replay}&\textbf{Total} \\
    \hline
    \hline
    \multirow{3}{*}{\textbf{ASVspoof2017 V2}}&Train&1,507&1,507&3,014\\  
    \cline{2-5}  
    &Development&760&950&1,710\\  
    \cline{2-5}  
    &Evaluation&1,298&12,008&13,306\\  
    \cline{2-5}
    \hline
    \hline
    \multirow{3}{*}{\textbf{ASVspoof2019 PA}}&Train&5,400&48,600&54,000\\
    \cline{2-5}
    &Development&5,400&24,300&29,700\\
    \cline{2-5}
    &Evaluation&18,090&116,640&134,730\\
    \cline{2-5}
    \hline
    \hline
    \multirow{4}{*}{\textbf{ASVspoof2021 PA}}&Evaluation&94,068&627,264&721,332\\
    \cline{2-5}
    &Progress&14,472&72,576&87,048\\
    \cline{2-5}
    &Hidden track 1&9,045&58,320&67,365\\
    \cline{2-5}
    &Hidden track 2&9,045&58,320&67,365\\
    \cline{2-5}
    \hline
  \end{tabular}
      }
\end{table}

\subsection{Experimental setup}
In our studies, the parameters in GFT are set according to our previous work in~\cite{GFCC_odyssey} to extract the GFCC feature. The proposed GFDCC and GFLDC features are then computed using the linear transformation parameters learned by ASVspoof 2017 V2 train set data as discussed in the previous section. We also applied cepstral mean and variance normalization (CMVN) on the extracted features to remove nuisance channel effects~\cite{Furui1981}.

We used two different classifiers for our studies. For the studies on ASVspoof 2017 V2 corpus, we used 512 mixture component based GMM models for genuine and spoofed speech following the baselines of that edition. On the other hand, we used one of the state-of-the-art systems LCNN as a classifier for the studies on ASVspoof 2019 PA. The LCNN architecture used in our studies follows that reported in \cite{karita21_interspeech}. It considers Adam optimizer with \(\beta_1\) = 0.9, \(\beta_2\) = 0.999, \(\epsilon = 10^{-8}\) \cite{https://1412.6980}. The initial learning rate of \(3 \times 10^{-4}\) is multiplied by 0.5 for every ten epochs, and the batch size is 8.

\section{Results and Analysis}
\label{section:Results and Analysis}
\subsection{Studies on ASVspoof 2017 V2}

We are first interested to explore the proposed three features on ASVspoof 2017 V2 corpus, which is a realistic replay database. We also consider enhanced CQCC baseline results reported in~\cite{ASVspoofV2} and our previous GFCC feature based representation for comparison which are shown in Table~\ref{tab:result1}. It is observed that the proposed GFLC outperforms our previous GFCC representation under all conditions, which proves the efficiency of the logarithm operation. In addition, our proposed device features GFDCC and GFLDC achieve the best results with and without CMVN, respectively. Again, CMVN is found to have a significant impact to boost the performance for all the front-ends, which is more evident when both train (T) and development (D) sets are used to learn the model. The use logE helps for CQCC feature when CMVN is applied as reported in~\cite{ASVspoofV2}, but it does not show such a trend from our studies on GFCC. Hence, we did not consider logE in case of proposed thress features. 

\begin{table}[t]
\normalsize
  \caption{Performance in EER (\%) for CQCC baseline, GFCC and the proposed GFLC, GFDCC and GFLDC with and without CMVN on ASVspoof 2017 V2 evaluation set, while model is either trained on train set (T) or train and development sets (T+D). logE stands for log Energy.} 
  \label{tab:result1}
  \centering
  \resizebox{\columnwidth}{!}
  {\begin{tabular}{|c|c|c|c|c|}
    \hline
    \multirow{2}{*}{\textbf{Feature}}
    &\multicolumn{2}{c|}{\textbf{without CMVN}}&\multicolumn{2}{c|}{\textbf{CMVN}}\\
    \cline{2-5}
    &\textbf{T}&\textbf{T+D}&\textbf{T}&\textbf{T+D} \\
    \hline
    CQCC \cite{ASVspoofV2}&30.79&23.97&19.74&15.33 \\
    CQCC plus logE \cite{ASVspoofV2}&34.95&29.31&13.74&12.24 \\
    \hline
    \hline
    GFCC&39.93&33.45&12.33&10.96 \\
    GFCC plus logE&40.70&32.37&12.70&11.04 \\
    \hline
    \hline
    GFLC without parallel FA&38.71&32.64&10.47&9.02\\
    GFDCC with parallel FA&\textbf{15.28}&\textbf{13.41}&11.93&10.63 \\
    GFLDC with parallel FA&15.88&13.79&\textbf{10.25}&\textbf{8.90} \\
    \hline
  \end{tabular}}
\end{table}

We now compare the proposed GFLC, GFDCC and GFLDC features based systems with some of the known single systems without data augmentation on ASVspoof 2017 V2 evaluation set, when the models are trained using combining the train and development set. Table~\ref{tab:result2} shows this comparison, where CQCC+GMM was the baseline for ASVspoof 2017 edition. Among all the systems compared in Table~\ref{tab:result2}, the proposed system GFLDC+GMM achieves the best performance. The CDOC, LFDCC and MFDCC are some other front-ends that considered device information during their extraction but performs inferior to our proposed device information feature GFDCC and GFLDC. Overall, the three features emerge as the strong front-end representations for the detection of replay attacks from the studies on ASVspoof 2017 V2 corpus. 

\begin{table}[t]
\normalsize
  \caption{Performance comparison in EER (\%) with some known single systems on ASVspoof 2017 V2 evaluation set.}
  \label{tab:result2}
  \centering
  \begin{tabular}{ll}
    \hline
    \textbf{System}&\textbf{EER}\\
    \hline
    CQCC + GMM (Baseline) \cite{ASVspoofV2}&15.33\\
    CQCC plus logE + i-vector \cite{ASVspoofV2}&12.93 \\
    FFT + ResNeWt \cite{9023158}&15.60 \\
    CQT + ResNeWt \cite{9023158}&12.21 \\
    FFT + LCNN\cite{Wang2019}&12.39\\
    qDFTspec + GMM\cite{5545402}&11.19\\
    ResNet-RV + OC-Softmax \cite{lou22_interspeech}&10.38 \\
    CDOC + DNN \cite{Longting_deviceFeat}&11.63 \\
    LFDCC + ResNet \cite{Longting_deviceFeat}&14.89 \\
    MFDCC + ResNet \cite{Longting_deviceFeat}&18.20 \\
    \hline
    GFLC + GMM&9.02 \\
    GFDCC + GMM&10.63 \\
    GFLDC + GMM&\textbf{8.90} \\
    \hline
  \end{tabular}
\end{table}

\subsection{Studies on ASVspoof 2019 PA}

In this subsection, we evaluate the performance of the proposed three features on ASVspoof 2019 PA database, which is a simulated replay attack corpus, and also compare them to some of the existing well-performing single systems including the two challenge baselines. Table~\ref{tab:result3} reports the results for this study, which reveals that the proposed three features perform better than most of the single systems, which is more evident for GFLC. Among all the single systems compared, CQT-MMPS based representation performs the best, however, it is noted that it is a hybrid feature representation that considers phase information along with the magnitude spectrum component. We would also like to highlight that the device-related linear transformation parameters for this study were still learned using the ASVspoof 2017 V2 training set, which is a realistic replay dataset and thus its characteristics are different from ASVspoof 2019 PA. However, the proposed three features are still able to get benefit from this device-related information as it performs better than the original GFCC front-end on ASVspoof 2019 PA showing the significance of the proposed approach. 


\begin{table}[t]
  \caption{Performance comparison in min t-DCF and EER (\%) with some known single systems without data augmentation on ASVspoof 2019 PA evaluation set. It is noted that the ASVspoof 2017 V2 dataset was not used in the the development of compared systems.}
  \label{tab:result3}
  \centering
  \begin{tabular}{lll}
    \hline
    \textbf{System}& \textbf{min t-DCF}&\textbf{EER}\\
    \hline
    CQCC + GMM (Baseline 1) \cite{ASVsppof2019_paper}&0.2454&11.04 \\
    LFCC + GMM (Baseline 2) \cite{ASVsppof2019_paper}&0.3017&13.54 \\
    LFCC + LCNN \cite{lavrentyeva19_interspeech}&0.1053&4.60 \\
    CQT + LCNN \cite{lavrentyeva19_interspeech}&0.0295&1.23 \\
    DCT + LCNN \cite{lavrentyeva19_interspeech}&0.0560&2.06 \\
    (LC-GRNN) + Softmax\cite{9360468}&0.0614&2.23\\
    (CQT-MMPS) + LCNN \cite{9360468}&\textbf{0.0240}&\textbf{0.90} \\
    CQCC + ResNet \cite{9023158}&0.0501&1.98 \\
    CQCC + ResNeWt \cite{9023158}&0.0419&1.67 \\
    GFCC + LCNN \cite{GFCC_odyssey}&0.0429&1.51 \\
    \hline
    GFLC + LCNN&0.0335&1.18 \\
    GFDCC + LCNN&0.0409&1.42 \\
    GFLDC + LCNN&0.0445&1.55 \\
    \hline
  \end{tabular}
\end{table}

\begin{table*}[t]
  \caption{Performance comparison in min t-DCF and EER (\%) without data augmentation, trained on ASVspoof 2017 V2 training set and evaluated on ASVspoof 2021 PA evaluation, progress,  hidden track 1 and hidden track 2 sets, respectively.}
  \label{tab:result4}
  \centering
  \begin{tabular}{lllllllll}
    \hline
    \multirow{2}{*}{\textbf{System}}
    &\multicolumn{2}{c}{\textbf{Evaluation}} &\multicolumn{2}{c}{\textbf{Progress}}&\multicolumn{2}{c}{\textbf{Hidden track 1}}&\multicolumn{2}{c}{\textbf{Hidden track 2}}\\
    \cmidrule(r){2-3}
    \cmidrule(r){4-5}
    \cmidrule(r){6-7}
    \cmidrule(r){8-9}
    & \textbf{min t-DCF}&\textbf{EER} & \textbf{min t-DCF}&\textbf{EER} & \textbf{min t-DCF}&\textbf{EER} & \textbf{min t-DCF}&\textbf{EER} \\
    \hline
    LFCC + GMM &0.9941&42.10&0.9539&40.56&0.9869&50.28&0.9823&43.43\\
    LFDCC + GMM &0.9297&35.36&0.9115&34.78&0.9223&33.33&\textbf{0.7267}&25.43\\
    MFCC + GMM &0.9446&38.23&0.9385&37.26&0.9420&38.87&0.8939&33.58\\
    MFDCC + GMM &0.9748&38.85&0.9816&38.97&0.8161&28.66&0.7852&28.25\\
    GFCC + GMM &0.9635&38.61&0.8981&35.56&0.9670&49.00&0.9079&36.72\\
    \hline
    GFLC + GMM &0.9506&37.18&0.8815&33.88&0.9643&49.83&0.8859&33.22\\
    GFDCC + GMM &\textbf{0.9154}&35.13&\textbf{0.8730}&32.93&\textbf{0.7147}&\textbf{23.28}&0.7509&26.36\\
    GFLDC + GMM &0.9196&\textbf{35.12}&0.8785&\textbf{32.84}&0.7216&23.48&0.7377&\textbf{25.17}\\
    \hline
  \end{tabular}
  \vspace{-3mm}
\end{table*}

\begin{table}[t]
  \caption{Performance comparison in min t-DCF and EER (\%) with some known single systems without data augmentation on ASVspoof 2021 PA evaluation set, trained on ASVspoof 2017 V2 training set or ASVspoof 2019 PA training set (T). It is noted that the ASVspoof 2017 V2 dataset was not used in development of the compared systems.}
  \label{tab:result5}
  \centering
  \begin{tabular}{llll}
    \hline
    \textbf{System}& \textbf{T} & \textbf{min t-DCF}&\textbf{EER}\\
    \hline
    CQCC + GMM (Baseline1)\cite{yamagishi:hal-03360794}&19&0.9434&38.07\\
    LFCC + GMM (Baseline2)\cite{yamagishi:hal-03360794}&19&0.9724&39.54 \\
    LFCC + LCNN (Baseline3)\cite{yamagishi:hal-03360794}&19&0.9958&44.77 \\
    RawNet2 (Baseline4)\cite{yamagishi:hal-03360794}&19&0.9997&48.60 \\
    World + GMM\cite{wang21_asvspoof}&19&\textbf{0.6853}&\textbf{24.88} \\
    SCMC + TDNN\cite{caceres21_asvspoof}&19&0.9339&36.10 \\
    CDOC + DNN\cite{Longting_deviceFeat}&19&0.9999&43.61 \\
    LFDCC + DNN\cite{Longting_deviceFeat}&19&0.9729&39.37 \\
    MFDCC + DNN\cite{Longting_deviceFeat}&19&0.9503&35.61 \\
    (Energy+FM) + ResNet\cite{RN170}&19&0.8836&35.26 \\
    GFCC + LCNN&19&0.9991&45.12 \\
    \hline
    GFLC + LCNN&19&0.9627&40.63 \\
    GFDCC + LCNN&19&0.9988&44.80 \\
    GFLDC + LCNN&19&1.0000&45.86 \\
    \hline
    CDOC + DNN \cite{Longting_deviceFeat}&17&0.9860&40.15 \\
    CDOC + ResNet \cite{Longting_deviceFeat}&17&0.9835&41.29 \\
    LFDCC + DNN \cite{Longting_deviceFeat}&17&\textbf{0.9149}&35.72 \\
    LFDCC + ResNet \cite{Longting_deviceFeat}&17&0.9443&37.97 \\
    MFDCC + DNN \cite{Longting_deviceFeat}&17&0.9999&38.84 \\
    MFDCC + ResNet \cite{Longting_deviceFeat}&17&0.9999&42.15 \\
    GFCC + GMM&17&0.9635&38.61 \\
    \hline
    GFLC + GMM&17&0.9506&37.18 \\
    GFDCC + GMM&17&0.9154&35.13 \\
    GFLDC + GMM&17&0.9196&\textbf{35.12} \\
    \hline
  \end{tabular}
  \vspace{-2mm}
\end{table}

\subsection{Studies on ASVspoof 2021 PA}

In this subsection, we assess the efficacy of the proposed three features on the ASVspoof 2021 PA database, a corpus involving realistic replay attacks. We also conduct a comparative analysis against several established single systems.

For the first set of studies, we used ASVspoof 2017 V2 training set to train the models and then subsequently evaluated on the four subsets of ASVspoof 2021 PA. The results are presented in Table~\ref{tab:result4}. GMM is chosen as the back-end classifier, and LFCC, linear frequency device cepstral coefficients (LFDCC), MFCC, mel frequency device cepstral coefficient (MFDCC), and GFCC are selected as comparison features in juxtaposition to the proposed GFLC, GFDCC and GFLDC. It is noted that the LFDCC and MFDCC are referenced methods in \cite{Longting_deviceFeat}. Table~\ref{tab:result4} reveals that GFDCC+GMM and GFLDC+GMM systems outperform the other six systems on most subsets.  Additionally, in comparison with GFCC+GMM system, GFDCC+GMM system exhibits a decrease in EER by 9.01\%, 7.40\%, 52.4\% and 28.21\%, respectively, across the four datasets. The minimum t-DCF also shows reductions of 5.00\%, 2.79\%, 26.09\% and 13.51\%, respectively. Compared with GFLC+GMM system, the performance of GFLDC+GMM on the four subsets has also been improved to varying degrees. It is noteworthy that previous experiments often neglected hidden track sets, making it challenging to obtain sufficient experimental results. Table~\ref{tab:result4} indicates that features utilizing device-related linear transformation significantly outperform original features, suggesting that this transformation better captures the distinctions between genuine and spoofed speech, particularly in retaining speech intervals where recording and replay devices exert a more pronounced influence on speech segments.



We now evaluate the performance of the proposed three features on ASVspoof 2021 PA using ASVspoof 2019 PA as the training set following the ASVspoof 2021 challenge protocol, which is designed under a domain mismatch condition of evaluating detection of realistic replay attacks when models are trained using simulated training set. Table~\ref{tab:result5} presents a comparison of proposed systems with some well-performing single systems. Among the single systems utilizing ASVspoof 2019 PA as the training set, world+GMM achieves the best results. It employs a world-vocoder-based replay channel estimation feature with log-spectrogram input and applies one-class and utterance-level GMM models. Systems incorporating constant-Q device octave coefficient (CDOC), LFDCC, and MFDCC as front-end features are based on device-related linear transformation reported in \cite{Longting_deviceFeat}. We find that the GFCC+LCNN and proposed three systems exhibit suboptimal performance when ASVspoof 2019 PA is used as the training set due to the simulated nature of this replay attack corpus. In contrast, when ASVspoof 2017 V2 is used as the training set, GFDCC+GMM and GFLDC+GMM surpass other known single systems, affirming the promising nature of GFDCC and GFLDC as front-end feature representations in scenarios where training and evaluation sets do not have a severe domain mismatch. In the future, we plan to explore severe domain mismatch variability for our feature representations to make them even effective for various conditions.

\section{Conclusions}
\label{section:Conclusions}
This work proposes three novel graph domain feature, namely GFLC that introduces logarithmic magnitude processing on GFT, as well as the GFDCC and GFLDC, which are derived using a device-related linear transformation on GFCC or GFLC feature, respectively. The device-related linear transformation parameters are computed from parallel train data from ASVspoof 2017 V2 aligned by DTW. We studied the proposed feature on ASVspoof 2017 V2, ASVspoof 2019 PA and ASVspoof 2021 PA datasets using GMM and LCNN as back-end classifiers. The studies on these databases demonstrated that the proposed three features outperform not only the baseline or GFCC feature, but also many state-of-the-art front-ends showing its robustness against replay attack detection due to the introduction of logarithmic processing as well as the incorporation of device information during their extraction. Lastly, the proposed method is highly dependent on availability of parallel training data and the future work will focus on finding ways to capture the device information without involvement of parallel training data. 


\section{Acknowledgement}
This work is supported by the National Natural Science Foundation of China under Grant 62001100.

\balance
\bibliographystyle{IEEEbib}
\bibliography{mybib}


%

\end{document}